\documentstyle[aps,epsf,twocolumn]{revtex}

\newcommand{\boldvec}[1]{\mbox{\boldmath$#1$}}
\newcommand{\smallvec}[1]{\mbox{\boldmath$\scriptstyle#1$}}
\newcommand{\annhilate}[3]{{#1}^{\phantom\dag}_{{#2}{#3}}}
\newcommand{\create}[3]{{#1}^{\dag}_{{#2}{#3}}}

\begin{document}

\draft

\title{Resonance superfluidity in a quantum degenerate Fermi gas}

\author{M. Holland$^1$, S.J.J.M.F. Kokkelmans$^1$, M.L. Chiofalo$^2$,
  and R.  Walser$^1$}

\address{$^1$JILA, University of Colorado and National Institute of
  Standards and Technology, Boulder, Colorado 80309-0440\\
  $^2$INFM and Scuola Normale Superiore, Piazza dei Cavalieri 7,
  I-56126 Pisa, Italy}

\date{\today}

\wideabs{

\maketitle

\begin{abstract}
  We consider the superfluid phase transition that arises when a
  Feshbach resonance pairing occurs in a dilute Fermi gas. We apply
  our theory to consider a specific resonance in potassium ($^{40}$K),
  and find that for achievable experimental conditions, the transition
  to a superfluid phase is possible at the high critical temperature
  of about $0.5\,T_F$.  Observation of superfluidity in this regime
  would provide the opportunity to experimentally study the crossover
  from the superfluid phase of weakly-coupled fermions to the
  Bose-Einstein condensation of strongly-bound composite bosons.
\vspace*{-1cm}
\end{abstract}

}

The achievement of Bose-Einstein condensation in atomic
vapors~\cite{bec} has given great impetus to efforts to realize
superfluidity in dilute fermionic alkali gases. While conditions of
quantum degeneracy have been obtained in potassium
($^{40}$K)~\cite{dfg}, the lowest achievable temperatures to date have
been limited to around $0.2T_F$~\cite{holland}. Although this limit is
essentially technical in nature, it appears likely that it will be
necessary to utilize a strong pairing mechanism yielding superfluid
transition temperatures close to this value.

\begin{figure}[h]
\begin{center}\
  \epsfysize=60mm \epsfbox{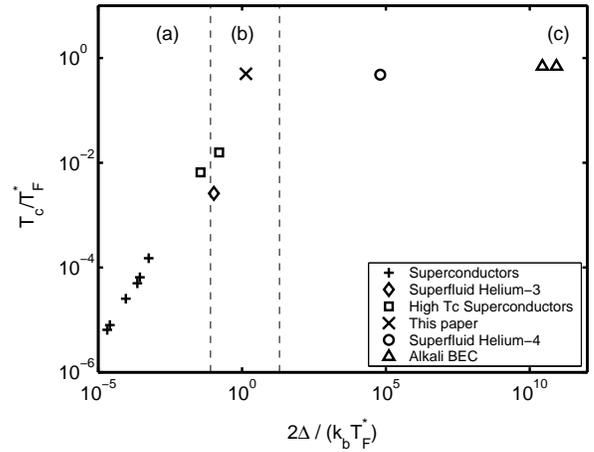}
\end{center}
\caption{
  A log-log plot showing six distinct regimes for quantum fluids. The
  transition temperature $T_c$ is shown as a function of the relevant
  gap energy~$2\Delta$.  Both quantities are normalized by an
  effective Fermi temperature $T_F^*$. For the BCS systems in region
  (a), and the systems in the cross-over region (b), $2\Delta$ is the
  energy needed to break up a fermion pair, and $T_F^*$ is the Fermi
  energy. For the systems in region (c), which are all strongly bound
  composite bosons and exhibit BEC phenomenology, $2\Delta$ is the
  smallest energy needed to break the composite boson up into two
  fermions, i.e. ionization to a charged atomic core and an electron,
  and $T_F^*$ is the ionic Fermi temperature.}
\label{fig1}
\end{figure}

Even in high-$T_c$ superconductors, the typical critical temperatures
are of the order of $10^{-2}T_F$.  In the context of strong-coupling
superconductivity there has been much work on constructing minimal
models to study the crossover from the seminal BCS theory~\cite{bcs}
for conventional superconductivity to the Bose-Einstein condensation
of tightly bound pairs, passing through non-perturbative regimes in
$T_c/T_F$~\cite{leggett,crossover}. In this letter, we treat
explicitly a short range quasibound resonant state by extending the
theory given in Refs.~\cite{friedberglee} to predict the existence of
a Feshbach resonance superfluidity in a gas of fermionic potassium
atoms. This system has an ultrahigh critical phase transition
temperature in close proximity to the Fermi temperature.  This is a
novel regime for quantum fluids, as illustrated in Fig.~\ref{fig1}
where our system and others which exhibit superfluidity or BEC are
compared.

The seminal Bardeen-Cooper-Schrieffer (BCS) theory~\cite{bcs} of
superconductivity applied to a dilute gas considers binary
interactions between particles in two distinguishable quantum states,
say $|\!\uparrow\rangle$ and $|\!\downarrow\rangle$. For a uniform
system, the fermionic field operators may be Fourier-expanded in a box
with periodic boundary conditions giving wavevector-$\boldvec{k}$
dependent creation and annihilation operators
$a^{\dag}_{\smallvec{k}\sigma}$ and $a_{\smallvec{k}\sigma}$ for
states $|\sigma\rangle$. At low energy, the binary scattering
processes are assumed to be completely characterized by the $s$-wave
scattering length $a$ in terms of a contact quasipotential
$U=4\pi\hbar^2an/m$, where $n$ is the number density. The Hamiltonian
describing such a system is given by
\begin{equation}
  H=\sum_{\smallvec{k}}\epsilon_{k}
  \bigl(\create{a}{\smallvec{k}}{\uparrow}
  \annhilate{a}{\smallvec{k}}{\uparrow} +
  \create{a}{\smallvec{k}}{\downarrow}
  \annhilate{a}{\smallvec{k}}{\downarrow} \bigr) 
  + U\sum_{\smallvec{k}_1\ldots\smallvec{k}_3}
  \create{a}{\smallvec{k}_1}{\uparrow}
  \create{a}{\smallvec{k}_2}{\downarrow}
  \annhilate{a}{\smallvec{k}_3}{\downarrow}
  \annhilate{a}{\smallvec{k}_4}{\uparrow},
\label{bcshamiltonian}
\end{equation}
where $\epsilon_k=\hbar^2k^2/2m$ is the kinetic energy, $m$ is the
mass, and the constraint
$\boldvec{k}_4=\boldvec{k}_1+\boldvec{k}_2-\boldvec{k}_3$ gives
momentum conservation. 

For a negative scattering length, the thermodynamic properties of the
gas show a superfluid phase transition at a critical temperature~$T_c$
which arises due to an instability towards the formation of
Cooper-pairs.  When the gas is dilute, as characterized by the
inequality $n|a|^3\ll1$ (or equivalently $k_F|a|\ll1$ where $k_F$ is
the Fermi wavenumber), the application of mean-field theory gives a
well-known solution for the ratio of $T_c$ to the Fermi
temperature~$T_F$~\cite{leggett}:
\begin{equation}
  \frac{T_c}{T_F}\sim\exp(-\frac{\pi}{2|a|k_F}).
\label{ratio}
\end{equation}
The exact prefactor to the exponential depends on the precise form of
the analytic integral approximations made in the derivation.  Several
papers have pointed out that the presence of a scattering resonance in
dilute alkali gases can be used to obtain a very large negative value
for the scattering length~\cite{stoof}. This promises the opportunity
for the system to enter the high-$T_c$ superfluidity regime as the
ratio in Eq.~(\ref{ratio}) approaches unity. However, direct
application of the BCS theory close to resonance then becomes
speculative due to the potential breakdown of a number of underlying
assumptions: (1)
Exactly on resonance the theory fails as the scattering length
  passes through $\pm\infty$ and the Hamiltonian in
  Eq.~(\ref{bcshamiltonian}) can not be defined,
(2)
For the mean field approach to be accurate it is required that
  there be many particles inside a volume associated with the spatial
  scale of a Cooper-pair. This condition begins to break down as $T_c$
  approaches $T_F$,
(3)
The theory of the dilute gas is formulated on a perturbation
  approach based on an expansion in the small parameter $n|a|^3$. When
  this parameter approaches unity the perturbation theory fails to
  converge.
These points show that care should be taken in applying
Eq.~(\ref{ratio}) near the point of resonance where the basis for the
conventional mean-field theory is not well-founded.

Despite these limitations, on general grounds, one would expect to be
able to derive a renormalizable low-energy effective field theory even
in close proximity to a resonance. This statement is based on the
identification that at relevant densities the range of the
interparticle potential is always orders of magnitude smaller than the
interparticle spacing. Here we present a theory of superfluidity in a
gas of dilute fermionic atoms which handles correctly the scattering
resonance and places the transition temperature to the superfluid
state in the experimentally accessible range.

While the scattering length $a$ usually characterizes the range of the
interatomic potential for a collision, this is a poor approximation in
the vicinity of a scattering resonance. The scattering properties are
completely determined by the positions of the bound states in the
interaction potentials. In a multichannel system, a bound state may
cross the threshold as a function of magnetic field and enter the
continuum, resulting in a field-dependent Feshbach scattering
resonance~\cite{feshbachVerhaar}.  As this occurs, the scattering
length becomes strongly dependent on the field, and exactly at
threshold it changes sign by passing through $\pm\infty$.

When such resonance processes occur, it is necessary to formulate the
Hamiltonian by separating out the resonance state and treating it
explicitly. This is motivated by the microscopic identification of two
types of scattering contributions: one from the scattering resonance,
and one from the background non-resonant processes that includes the
contributions from all the other bound states. The non-resonant
contributions give rise to a background scattering length $a_{\rm bg}$
which is a good characterization of the potential range. The
corresponding quasipotential in that case is given by $U_{\rm
  bg}=4\pi\hbar^2a_{\rm bg}n/m$.  The Feshbach resonance occurs due to
a coupling with a molecular state, that is long-lived in comparison
with characteristic non-resonant collision timescales. This state is a
composite boson which is described by bosonic annihilation operators
$\annhilate{b}{\smallvec{k}}{}$. It is parameterized by a detuning
energy from threshold, denoted by $2\nu$, that is dependent on the
value of the magnetic field. The coupling strength of
$\annhilate{b}{\smallvec{k}}{}$ to the two-particle continuum is well
characterized by a single coupling constant $g$, independent of
$\boldvec{k}$. These considerations imply that the Hamiltonian given
in Eq.~(\ref{bcshamiltonian}) is not sufficient to account for the
important resonance processes and must be extended to incorporate
explicitly the coupling between the atomic and molecular gases:
\begin{eqnarray}
  H&=&2\nu\sum_{\smallvec{k}}\create{b}{\smallvec{k}}{}
  \annhilate{b}{\smallvec{k}}{} + \sum_{\smallvec{k}}\epsilon_{k}
  \bigl(\create{a}{\smallvec{k}}{\uparrow}
  \annhilate{a}{\smallvec{k}}{\uparrow} +
  \create{a}{\smallvec{k}}{\downarrow}
  \annhilate{a}{\smallvec{k}}{\downarrow} \bigr) \nonumber\\
  &&\quad{}+ U_{\rm bg}\sum_{\smallvec{k}_1\ldots\smallvec{k}_3}
  \create{a}{\smallvec{k}_1}{\uparrow}
  \create{a}{\smallvec{k}_2}{\downarrow}
  \annhilate{a}{\smallvec{k}_3}{\downarrow}
  \annhilate{a}{\smallvec{k}_4}{\uparrow} \nonumber\\
  &&\quad{}+ g\sum_{\smallvec{k},\smallvec{q}}
  \create{b}{\smallvec{q}}{}
  \annhilate{a}{\frac{\smallvec{q}}{2}+\smallvec{k}}{\uparrow}
  \annhilate{a}{\frac{\smallvec{q}}{2}-\smallvec{k}}{\downarrow} +
  \annhilate{b}{\smallvec{q}}{}
  \create{a}{\frac{\smallvec{q}}{2}-\smallvec{k}}{\downarrow}
  \create{a}{\frac{\smallvec{q}}{2}+\smallvec{k}}{\uparrow}.
\end{eqnarray}
Evolution generated by this Hamiltonian conserves the particle number
$N=\sum_{\smallvec{k}}\bigl(\create{a}{\smallvec{k}}{\uparrow}
\annhilate{a}{\smallvec{k}}{\uparrow} +
\create{a}{\smallvec{k}}{\downarrow}
\annhilate{a}{\smallvec{k}}{\downarrow}
\bigr)+2\sum_{\smallvec{k}}\create{b}{\smallvec{k}}{}
\annhilate{b}{\smallvec{k}}{}$. Note that the Hamiltonian does not
contain $a$ explicitly, and that the field dependence of the
scattering is completely characterized by the parameters: $g$, $\nu$,
and~$U_{\rm bg}$. The magnitude of $g$ is derived in the following
way. We define $\kappa$ as the product of the magnetic field width of
the resonance and the magnetic moment difference of the Feshbach state
and the continuum state. For large values of $\nu$, the boson field
$\annhilate{b}{\smallvec{k}}{}$ can be adiabatically eliminated from
the theory, and then $g=\sqrt{\kappa U_{\rm bg}}$ is required in order
for the scattering properties to have the correct dependence on
magnetic field~\cite{couplingConstant}.

The essential point is that this Hamiltonian, founded on the
microscopic basis of resonance scattering, is well-behaved at all
detunings $\nu$; even for the pathological case of exact resonance.
The diluteness criterion is now given by constraints which require
both the potential range and the spatial extent of the Feshbach
resonance state, to be much smaller than the interparticle spacing
(e.g.\ $n|a_{\rm bg}|^3\ll1$). 

We apply this Hamiltonian to derive the self-consistent mean-fields
for given thermodynamic constraints by formulating a
Hartree-Fock-Bogoliubov theory~\cite{bosonmodel}. The mean-fields
present include the fermion number $f=\sum_{\smallvec{k}}
\langle\create{a}{\smallvec{k}}{\uparrow}
\annhilate{a}{\smallvec{k}}{\uparrow}\rangle$, the molecule field
$\phi_m=\langle \annhilate{b}{\smallvec{k}=0}{}\rangle$ taken to be a
classical field, and the pairing-field $p=\sum_{\smallvec{k}}
\langle\annhilate{a}{\smallvec{k}}{\uparrow}
\annhilate{a}{-\smallvec{k}}{\downarrow}\rangle$~\cite{magnetization}.
It is well-known that such a theory must be renormalized in order to
remove the ultraviolet divergence which arises from the incorporation
of second-order vacuum contributions.  This implies replacing the
physical parameters in the Hamiltonian, $U$, $g$, and $\nu$, by
renormalized values so that observables are independent of a high
momentum cut-off used in the formulation of the effective
field-theory~\cite{followingpaper}. In order to diagonalize the
Hamiltonian, we construct Bogoliubov quasiparticles according to the
general canonical transformation~\cite{bogval}
\begin{equation}
  \left(\begin{array}{c}
      \annhilate{\alpha}{\smallvec{k}}{\uparrow}\\
      \create{\alpha}{-\smallvec{k}}{\downarrow}
  \end{array}
\right)=\left(
\begin{array}{cc}
  \cos\theta & -e^{i\gamma}\sin\theta \\
  e^{i\gamma}\sin\theta & \cos\theta
\end{array}
\right)\left(
\begin{array}{c}
  \annhilate{a}{\smallvec{k}}{\uparrow}\\
  \create{a}{-\smallvec{k}}{\downarrow}
\end{array}
\right).
\end{equation}
Given single particle energies, $U_k=\epsilon_k-\mu+Uf$, where $\mu$
is the chemical potential, and the gap parameter in the quasiparticle
spectrum $\Delta=Up-g\phi_m$, the two transformation angles are
specified as $\tan(2\theta)=|\Delta|/U_k$ and
$\phi_m=|\phi_m|\exp(i\gamma)$. The corresponding quasiparticle
spectrum is $E_k=\sqrt{U_k^2+\Delta^2}$. Dropping terms of higher
order than quadratic in the fermion operators, gives the resulting
many-body Hamiltonian
\begin{eqnarray}
  H-\mu N&=&2(\nu-\mu)|\phi_m|^2\nonumber\\&&{}+
  \sum_{\smallvec{k}}\Bigl(U_k + E_k\bigl(
  \create{\alpha}{\smallvec{k}}{\uparrow}
  \annhilate{\alpha}{\smallvec{k}}{\uparrow}+
  \create{\alpha}{\smallvec{k}}{\downarrow}
  \annhilate{\alpha}{\smallvec{k}}{\downarrow}-1\bigr) \Bigr).
\label{manybody}
\end{eqnarray}
which is now in diagonal form.

The next task is to calculate the thermodynamic solutions. Equilibrium
populations for the quasiparticles are given by the Fermi-Dirac
distribution.  The fermion number and pairing field are not only
inputs to the Hamiltonian, but also determine the quasiparticle
spectrum. Therefore, they must be self-consistent with the values
derived by summing the relevant equilibrium density matrix elements
over all wave numbers. In practice, at a given temperature, chemical
potential, and molecule number $\phi_m$, this requires an iterative
method to locate self-consistent values for $f$ and $p$. The value of
$\phi_m$ is calculated by minimizing the grand potential
$\Phi_G=-k_bT\ln\Xi$ at fixed temperature and chemical potential, with
$k_b$ denoting Boltzmann's constant. The partition function
$\Xi=\mbox{Tr}[\exp(-(H-\mu N)/k_bT)]$ is found from
Eq.~(\ref{manybody}). This procedure is mathematically equivalent to
minimizing the Helmholtz free-energy at fixed temperature and density
and corresponds uniquely to the maximum entropy solution. This
solution has an associated particle number,
$\bigl<N\bigr>=-\partial\Phi_G/\partial\mu$ taken at constant
temperature and volume, which must match the actual particle density
of the gas, so that the final step is to adjust the chemical potential
until this condition is satisfied. The whole procedure is repeated
over a range of temperatures to determine the locus of thermodynamic
equilibrium points. For large positive detunings, where the molecule
field could be eliminated from the theory entirely, regular BCS theory
emerges. For this case, when the scattering length $a$ is negative the
behaviour of the critical temperature on $1/a$ is given by the usual
exponential law~\cite{leggett}.

\begin{figure}[h]
\begin{center}\
  \epsfysize=50mm \epsfbox{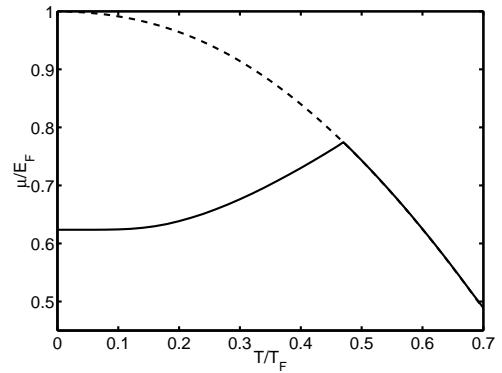}
\end{center}
\caption{
  Chemical potential as a function of temperature for the system of
  resonance pairing (solid line). The second order phase transition
  occurs at $T_c\approx 0.5\;T_F$ where a clear cusp is visible. The
  dashed line shows the chemical potential of a non-interacting Fermi
  gas.}
\label{fig2}
\end{figure}

\begin{figure}[h]
\begin{center}\
  \epsfysize=50mm \epsfbox{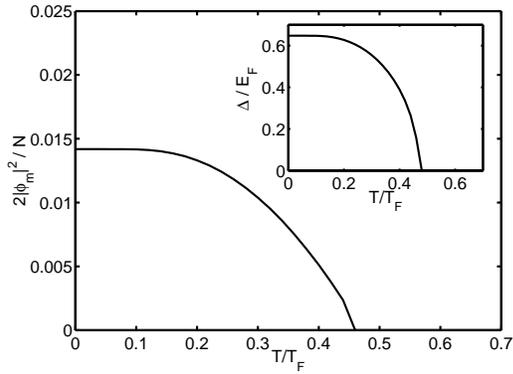}
\end{center}
\caption{
  The temperature at the phase transition is
  also visible from the amplitude of the molecular field. This
  amplitude is non-zero only when the broken symmetry exists in the
  region $T<T_c$.  For $T=0$, the molecules form a Bose condensed
  fraction of 1.5\% of the total gas sample. The inset shows the
  behaviour of the gap $\Delta=Up-g\phi_m$. The critical temperature
  $T_c$ can be related to the value of the gap at $T=0$. For
  comparison, in superconductors the analagous gap is simply the
  binding energy of a fermion pair.}
\label{fig3}
\end{figure}

As an example of the application of this theory, we study the
experimentally relevant system of fermionic $^{40}$K atoms equally
distributed between the two hyperfine states which have the lowest
internal energy in the presence of a magnetic field. The values of our
interaction parameters $a_{\rm{bg}}=176$ a$_{0}$ and $\kappa/k_b =657$
$\mu $K are obtained from~\cite{bohn}. We fix the total density to be
$n=10^{14}$~cm$^{-3}$, a typical experimental value expected for this
quantum degenerate gas in an optical trap.  We set the detuning to be
$\nu=+E_F$ so that the quasi-bound state is detuned slightly above the
atomic resonance. For a temperature above $T_c$, the grand potential
surface is shaped like a bowl, and the value of $\phi_m$ which
minimizes the grand potential is $\phi_m=0$, associated with the
self-consistent solution $p=0$. For $T<T_c$,the grand potential
surface is shaped like a Mexican hat, and its minimum is given by a
$\phi_m$ with non-zero amplitude and an undetermined phase. The
superfluid phase transition therefore leads to a spontaneously broken
symmetry. The value of $T_c$ can be clearly found from
Figs.~\ref{fig2} and~\ref{fig3}, where we show the chemical potential,
the molecular density, and the gap as a function of temperature. We
find for our parameter set for $^{40}$K and almost zero detuning a
remarkably high value for the critical temperature
$T_c\approx0.5\,T_F$, i.e.\ $T_c\approx0.6$~$\mu$K. Furthermore we
find a weak dependence of $T_c\approx0.5\,T_F$ on the density, so that
the value of $T_c$ has more-or-less the same density behavior as
$T_F$.  When we increase the detuning to $\nu=+17.6\,E_F$ (this
corresponds to a magnetic field detuning of $0.5$~G away from the
Feshbach resonance), the value of $T_c$ drops to approximately
$0.25\,T_F$.

The system of $^{40}$K atoms, equally distributed among the two lowest
hyperfine states, is a good candidate for demonstrating the superfluid
phase transition. It not only exhibits a Feshbach resonance, but also,
the inelastic binary collision events are energetically forbidden.
Three-body interactions are highly suppressed, since the asymptotic
three-body wave function should consist of a product of three $s$-wave
two-body scattering wave functions. In a three-body interaction,
two-particles are always in the same initial hyperfine state, and
therefore the corresponding $s$-wave state is forbidden. The only
three body relaxation could come from asymptotic $p$-waves, but these
have very little contribution at the low temperatures considered.
Although the detailed three-body collision problem is an intricate
one, this asymptotic statistical effect should lead to a large
suppression of the vibrational relaxation of quasi-bound molecules.

Current experimental techniques for ultracold gases do not produce
samples which are spatially uniform. An optical dipole trap may be
needed to confine the high field seeking atoms, and the conditions for
the superfluid phase transition would be satisfied first in the trap
center where the density is highest. The presence of the quasi-bound
molecules may be a very useful aspect allowing direct observation of
the phase transition through imaging the molecular field.

In conclusion, we have shown that resonance pairing in an alkali gas
yields a quantum fluid that can undergo a superfluid phase transition
at a temperature comparable to the Fermi temperature. This
extraordinary property places this system in a regime which lies in
between BCS-like superconductors, and bosonic systems which may
undergo BEC.  Since the transition temperature is larger than the
lowest temperatures already achieved in a degenerate Fermi gas, it
should be possible to study this new type of quantum matter and to
quantitatively compare with our predictions.

\section*{Acknowledgements}
We thank J. Cooper, E. Cornell, D. Jin, C. Wieman, and B. DeMarco for
very stimulating discussions. Support is acknowledged for M.H. from
the National Science Foundation, for S.K. from the U.S. Department of
Energy, Office of Basic Energy Sciences via the Chemical Sciences,
Geosciences and Biosciences Division, for M.C. from the INFM and SNS,
Pisa (Italy), and for R.W. from the APART fellowship, Austrian Academy
of Sciences.

{\em Note Added:} A similar treatment has recently been proposed by
Timmermans {\em et al.}~\cite{timmermans}.


\begin{references}
  
  \vspace*{-1.5cm}

\bibitem{bec} M. H. Anderson, J. R. Ensher, M. R. Matthews, C. E.
  Wieman, and E.  A. Cornell, Science {\bf 269}, 198 (1995); K. B.
  Davis, M.-O. Mewes, M. R. Andrews, N. J. van Druten, D. S. Durfee,
  D. M. Kurn, and W. Ketterle, Phys.\ Rev.\ Lett.\ {\bf 75}, 3969
  (1995); C. C. Bradley, C. A. Sackett, J. J. Tollett, and R. G.
  Hulet, Phys.\ Rev.\ Lett.\ {\bf 75}, 1687 (1995); {\bf 79}, 1170(E)
  (1997).
  
\bibitem{dfg} B. DeMarco and D. S. Jin, Science {\bf 285}, 1703
  (1999).
  
\bibitem{holland} M. J.  Holland, B. DeMarco, and D. S. Jin, Phys.\ 
  Rev.\ A {\bf 61}, 053610 (2000); B. DeMarco, S.B. Papp, and D.S.
  Jin, Phys. Rev. Lett. {\bf 86}, 5409 (2001).

\bibitem{bcs} J. Bardeen, L. N. Cooper, and J.~R.~Schrieffer, Phys.\ 
  Rev. {\bf 108}, 1175 (1957); J. R. Schrieffer, {\it Theory of
    Superconductivity}, Perseus Books, Reading, Massachusetts, (1999).
  
\bibitem{leggett} A. G. Leggett, J. Phys. (Paris) {\bf C7}, 19 (1980);
  M. Houbiers and H. T. C. Stoof, Phys.\ Rev.\ A {\bf 59}, 1556-1561
  (1999); G. Bruun, Y. Castin, R. Dum {\em et al.}, Eur.  Phys. J. D
  {\bf 7}, 433--439 (1999); H. Heiselberg, C. J. Pethick, H. Smith,
  and L. Viverit, Phys. Rev. Lett. {\bf 85}, 2418 (2000).
  
\bibitem{crossover} See M. Randeria and references therein in {\it
    Bose-Einstein condensation}, ed. by A.  Griffin, D.W. Snoke and S.
  Stringari, Cambridge Un. Press, Cambridge (1995).
  
\bibitem{friedberglee} J. Ranninger and S. Robaszkiewicz, Physica B
  {\bf 53}, 468 (1985); R. Friedberg and T. D. Lee, Phys. Rev. B {\bf
    40}, 6745 (1989).
  
\bibitem{stoof} H. T. C. Stoof, M. Houbiers, C. A. Sackett, and R. G. Hulet,
  Phys.\   Rev.\ Lett. {\bf 76}, 10 (1996); R. Combescot, Phys.
  Rev. Lett. {\bf 83}, 3766 (1999).

\bibitem{feshbachVerhaar} H. Feshbach, Ann. Phys. {\bf 5}, 357 (1958);
  E. Tiesinga, B. J. Verhaar, and H. T. C. Stoof, Phys. Rev. A {\bf
    47}, 4114 (1993); S. Inouye, M. R. Andrews, J. Stenger, H.-J.
  Miesner, D. M. Stamper-Kurn, and W. Ketterle, Nature {\bf 392}, 151
  (1998).
  
\bibitem{couplingConstant} This expression for $g$ is chosen so that
  $a$ obeys the correct field dependence. For further discussion see
  E. Timmermans {\em et al.}, Phys.\ Rep.\ {\bf 315} 199 (1999).
  
\bibitem{bosonmodel} An analagous field-theory is derived for a
  bosonic model in M. Holland, J. Park, and R. Walser, Phys. Rev.
  Lett., {\bf 86}, 1915 (2001).
  
\bibitem{magnetization} A magnetization field $\sum_{\smallvec{k}}
  \langle\create{a}{\smallvec{k}}{\uparrow}
  \annhilate{a}{\smallvec{k}}{\downarrow}\rangle$ is also included in
  our formulation. However, we drop this term in our discussion since
  it is identically zero in the spin-symmetric case considered here.
  Inclusion gives a slightly more general treatment, and requires the
  addition of a spin-rotation following the transformation to
  Bogoliubov quasiparticles.
  
\bibitem{followingpaper} S.J.J.M.F. Kokkelmans, R. Walser, M.
  Chiofalo, J. Milstein, and M. Holland, to be published.
  
\bibitem{bogval} N. N. Bogoliubov, Nuovo Cimento {\bf 7}, 6 (1958);
  {\bf 7}, 794 (1958); J. Valatin, Nuovo Cimento {\bf 7}, 843 (1958).
  
\bibitem{bohn} John L. Bohn, Phys. Rev. A {\bf 61}, 053409 (2000).
  
\bibitem{timmermans} E. Timmermans, K. Furuya, P. W. Milonni, A. K.
  Kerman, Phys.  Lett. A {\bf 285}, 228 (2001).
  
\end{references}
\end{document}